\documentclass[a4paper,11pt]{article}
\usepackage{pos}

\title{Origins and impacts of dynamical diquark correlations \\ -- A continuum Schwinger functional approach --}
\ShortTitle{Origins and impacts of dynamical diquark correlations}

\author*[a]{Jorge Segovia}

\affiliation[a]{Departamento de Sistemas F\'isicos, Qu\'imicos y Naturales, Universidad Pablo de Olavide, \\
Ctra. de Utrera km. 1, 41013 Sevilla, Spain}

\emailAdd{jsegovia@upo.es}

\abstract{This conference proceedings contribution emphasizes the emergent hadron mass paradigm, which accounts for the majority of the visible mass in the universe, beyond the Higgs boson mechanism. The study delves into the Landau gauge gluon propagator and the dynamical generation of gluon mass, as well as the dressed-quark propagator and dynamical chiral symmetry breaking. It also tackles the baryon bound state problem through the Poincaré-covariant Faddeev equation, analyzing the composition and masses of octet and decuplet baryons. The document concludes with a discussion on the electromagnetic form factors of the nucleon and its first radial excitation, providing insights into the quark-diquark structure within baryons.}

\FullConference{10th International Conference on Quarks and Nuclear Physics (QNP2024)\\
8-12 July, 2024\\
Barcelona, Spain\\}

\definecolor{olive}{rgb}{0.3, 0.4, .1}
\definecolor{fore}{RGB}{249,242,215}
\definecolor{back}{RGB}{51,51,51}
\definecolor{title}{RGB}{255,0,90}
\definecolor{dgreen}{rgb}{0.,0.6,0.}
\definecolor{gold}{rgb}{1.,0.84,0.}
\definecolor{JungleGreen}{cmyk}{0.99,0,0.52,0}
\definecolor{BlueGreen}{cmyk}{0.85,0,0.33,0}
\definecolor{RawSienna}{cmyk}{0,0.72,1.00,0.85}
\definecolor{Magenta}{cmyk}{0,1,0,0}
\definecolor{DarkerRed}{rgb}{1.00,0.10,0.10}
\definecolor{DarkerFuchsia}{RGB}{118,0,118}

\begin{document}
\maketitle


\section{Continuum Schwinger function method}

\begin{figure}[!t]
\begin{center}
\includegraphics[width=0.47\textwidth, height=0.25\textheight]
{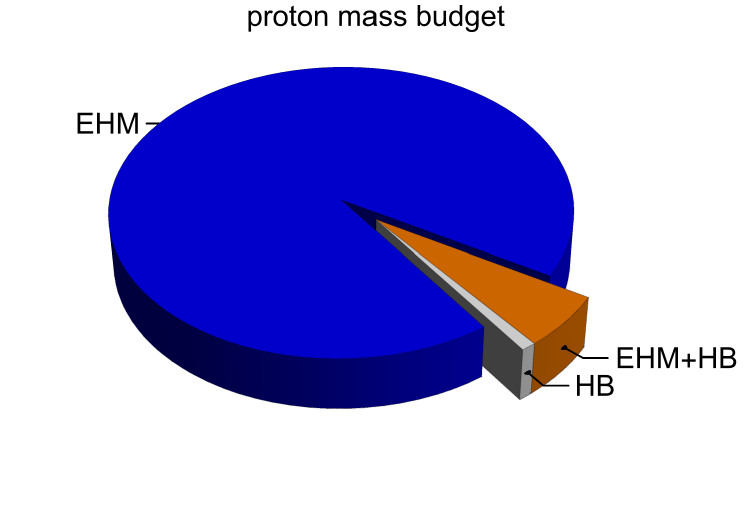}
\hspace*{0.25cm}
\includegraphics[width=0.47\textwidth, height=0.25\textheight]
{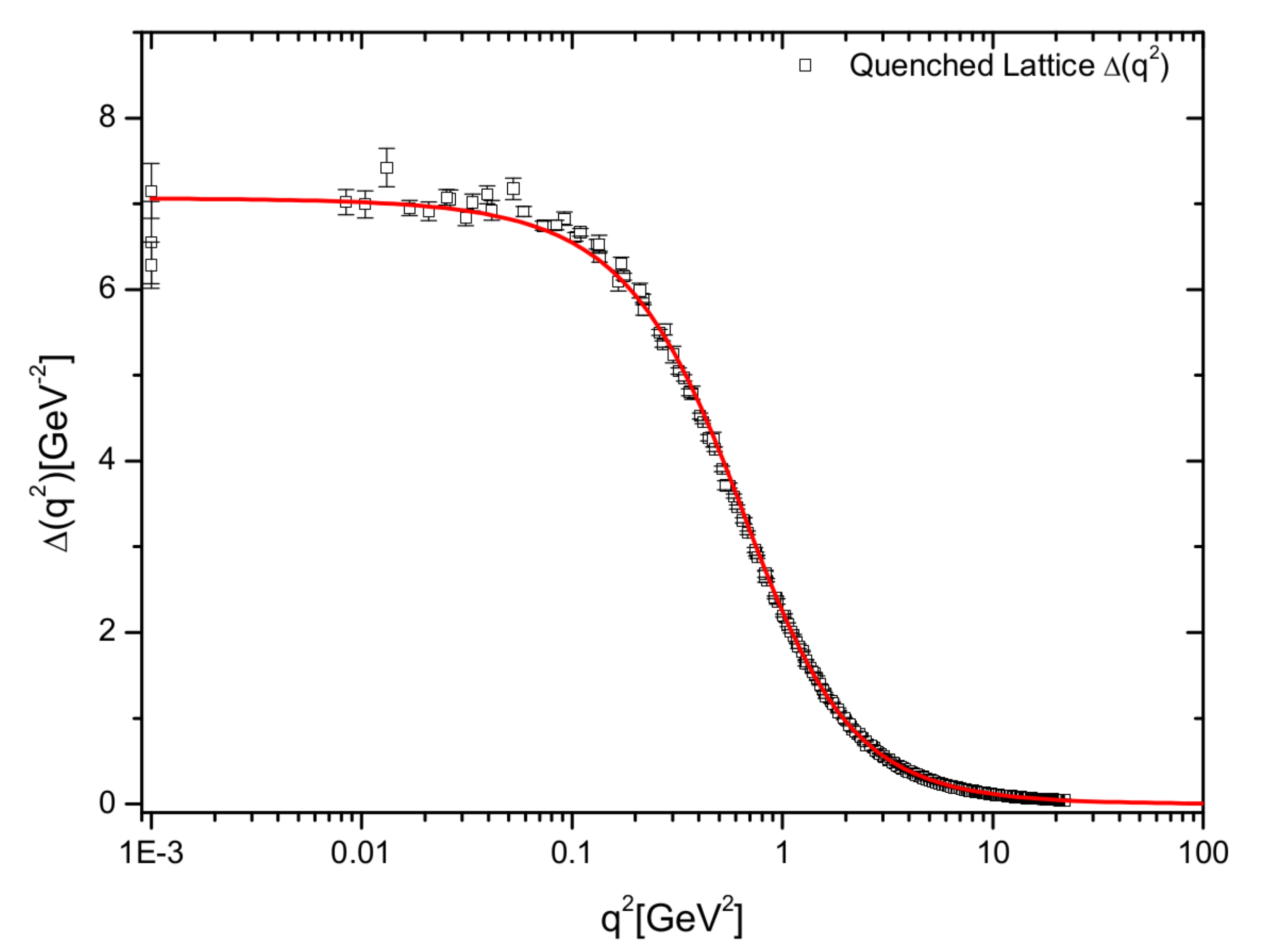}
\includegraphics[width=0.47\textwidth, height=0.25\textheight]
{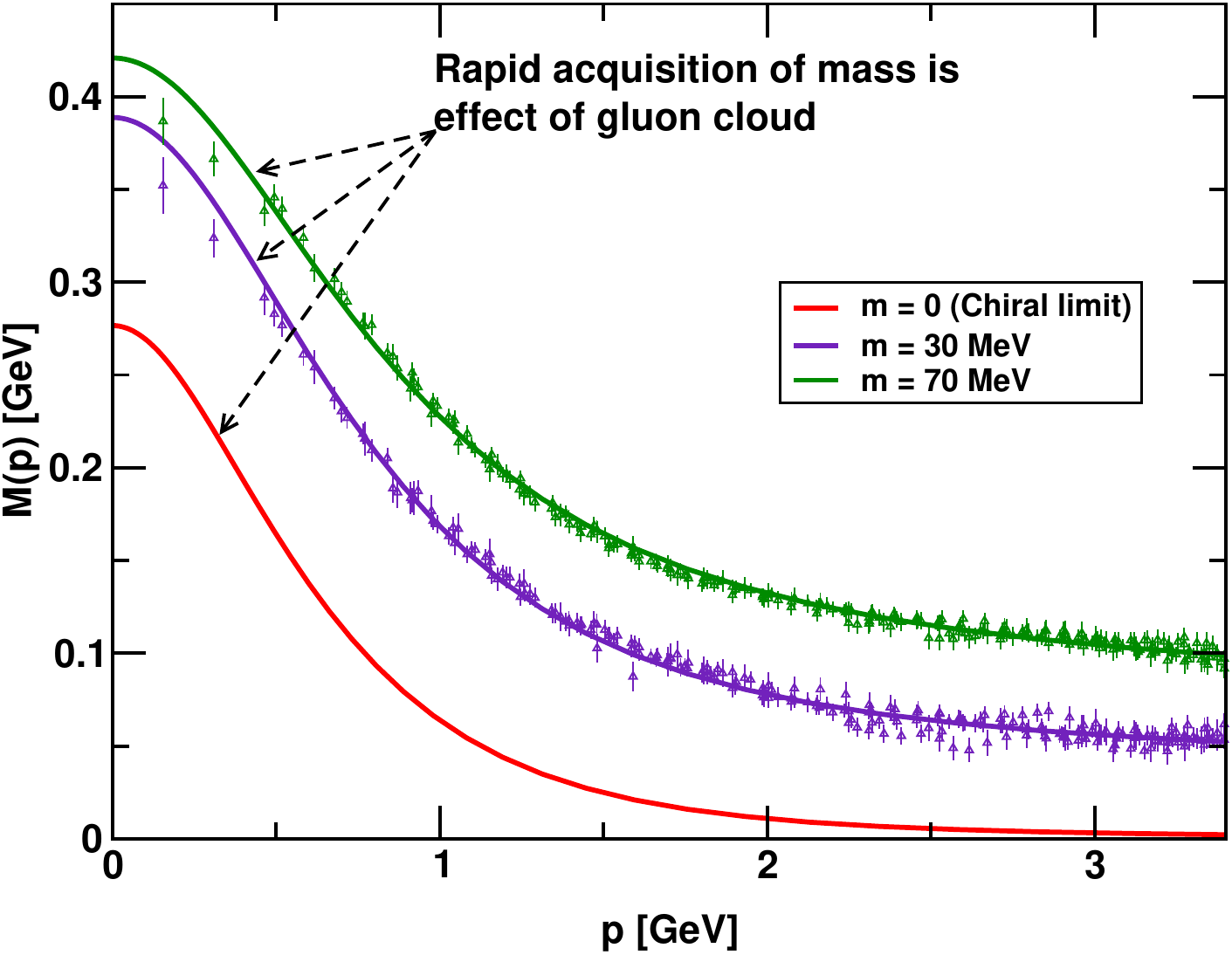}
\hspace*{0.25cm}
\includegraphics[width=0.47\textwidth, height=0.26\textheight]
{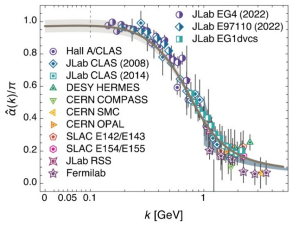}
\end{center}
\vspace*{-0.20cm}
\caption{\label{fig:QCD-DSEs}
\emph{Top-left panel:} Proton mass budget, drawn using a Poin\-car\'e-invariant decomposition: EHM $= 94$\%; Higgs boson (HB) contribution $= 1$\%; and EHM+HB interference $= 5$\%.
\emph{Top-right panel:} Solutions of the quark DSE (gap equation) for $M(p^2)$, obtained using the current-quark mass that best fit lattice-QCD results; and the (red) lowest solid curve represents the gap equation's solution in the chiral limit.
\emph{Bottom-left panel:} The quenched lattice gluon propagator $\Delta(q^2)$ and an adapted solution.
\emph{Bottom-right panel:} Process-independent effective charge, calculated by combining results from continuum and lattice studies of QCD's gauge sector.
}
\end{figure}

The Higgs boson (HB) mechanism~\cite{Englert:2014zpa, Higgs:2014aqa} is widely known as the mass-generating process within the Standard Model. However, this mechanism produces just 1\% of the Universe's visible mass which is constituted by nuclei, the mass of each such nucleus is basically the sum of the masses of the nucleons they contain, and only $9\,$MeV of a nucleon's mass, $m_N=940\,$MeV, is directly generated by the Higgs boson coupling (see top-left panel of Fig.~\ref{fig:QCD-DSEs}).

Nature has then another very effective mass-generating mechanism. Today, this is called emergent hadron mass (EHM)~\cite{Roberts:2021nhw, Binosi:2022djx, Papavassiliou:2022wrb, Ding:2022ows, Ferreira:2023fva, Carman:2023zke} and it basically collects three pillars within the continuum Schwinger function method (CSM): the running gluon and quark masses as well as the existence of a well-defined infrared process-independent effective strong coupling constant.

The gluon propagator in Landau gauge assumes the totally transverse form
\begin{equation}
i \Delta_{\mu\nu}(q) = -i P_{\mu\nu}(q) \Delta(q^2)\,, \quad\quad P_{\mu\nu}(q) = g_{\mu\nu} - q_\mu q_\nu / q^2 \,,
\end{equation}
where the scalar form factor $\Delta(q^2)$ is related to the all-order gluon self-energy. Lattice-QCD simulations~\cite{Bogolubsky:2009dc, Bowman:2007du, Cucchieri:2009zt} have recently confirmed~\cite{Cornwall:1981zr, Mandula:1987rh} that this form factor saturates in the deep infrared (see top-right panel of Fig.~\ref{fig:QCD-DSEs}) indicating gluon mass generation. A demonstration of how this occurs at the level of the gluon Dyson-Schwinger Equation (DSE) is given in Ref.~\cite{Aguilar:2008xm}. The transition from a massless to a massive gluon propagator can be implemented as $\Delta^{-1}(q^2) = q^2 J(q^2) + m^2(q^2)$, where $J(q^2)$ is the gluon's dressing function and $m^2(q^2)$ is the momentum dependent gluon mass, dynamically generated by the Schwinger mechanism~\cite{Schwinger:1962tn, Schwinger:1962tp}. This can be summarize as follows: the vacuum polarization of a gauge boson that is massless at the level of the original Lagrangian may develop a massless pole, whose residue can be identified with $m^2(0)$. The origin of the aforementioned poles is due to purely non-perturbative dynamics. They act as composite Nambu-Goldstone bosons which are colored, massless and have a longitudinal coupling. These features maintain gauge invariance and makes them disappear from any on-shell $S$-matrix element.

The dressed-quark propagator in Landau gauge can be written in the following form:
\begin{equation}
S_f(p) =  -i \gamma\cdot p\, \sigma_V^f(p^2) + \sigma_S^f(p^2) = 1/[i\gamma\cdot p\, A_f(p^2) + B_f(p^2)]\,.
\label{eq:Sq}
\end{equation}
It is known that for light-quarks the wave function renormalization, $Z_f(p^2)=1/A_f(p^2)$, and dressed-quark mass, $M_f(p^2)=B_f(p^2)/A_f(p^2)$, receive strong momentum-dependent corrections at infrared momenta~\cite{Bhagwat:2003vw}: $Z_f(p^2)$ is suppressed and $M_f(p^2)$ enhanced (see bottom-left panel of Fig.~\ref{fig:QCD-DSEs}). These features are an expression of dynamical chiral symmetry breaking (DCSB) which has long been argued to provide the key to understanding the pion, Nature's most fundamental Nambu-Goldstone boson, with its unusually low mass and structural peculiarities~\cite{Horn:2016rip, Aguilar:2019teb, Arrington:2021biu}.

The most up-to-date result for QCD's effective charge is drawn in the bottom-right panel of Fig.~\ref{fig:QCD-DSEs}. It was obtained~\cite{Cui:2019dwv} by combining contemporary results from continuum analyses of QCD's gauge sector and lattice-QCD configurations generated with three domain-wall fermions at the physical pion mass~\cite{RBC:2014ntl, Boyle:2015exm, Boyle:2017jwu, Zafeiropoulos:2019flq} to obtain a parameter-free prediction. The most notable features of the point-wise behavior of the effective charge are~\cite{Deur:2022msf}: (i) there is no Landau pole that afflicts perturbative QCD, (ii) this coupling saturates to a large, finite value at $Q^2=0$ recovering QCD's conformal character of the classical theory and (iii) it resolves the problem of Gribov copies. Significantly, too, $\hat\alpha(k)$ is practically identical to the effective charge $\alpha_{g_1}$, for which data is plentiful. Thus, with $\hat\alpha(k)$, one has in hand an excellent candidate for that long-sought running coupling which characterises QCD interactions at all momentum scales.

Once the three pillars that support the CSM paradigm of EHM has been sketched, phenomenologists and theoreticians are engaged in identifying their observable consequences. The challenge for experimentalists is to test and measure this body of predictions so that the boundaries of the Standard Model may finally be drawn~\cite{Proceedings:2020fyd}.


\begin{figure}[!t]
\begin{center}
\includegraphics[width=0.70\textwidth, height=0.15\textheight]{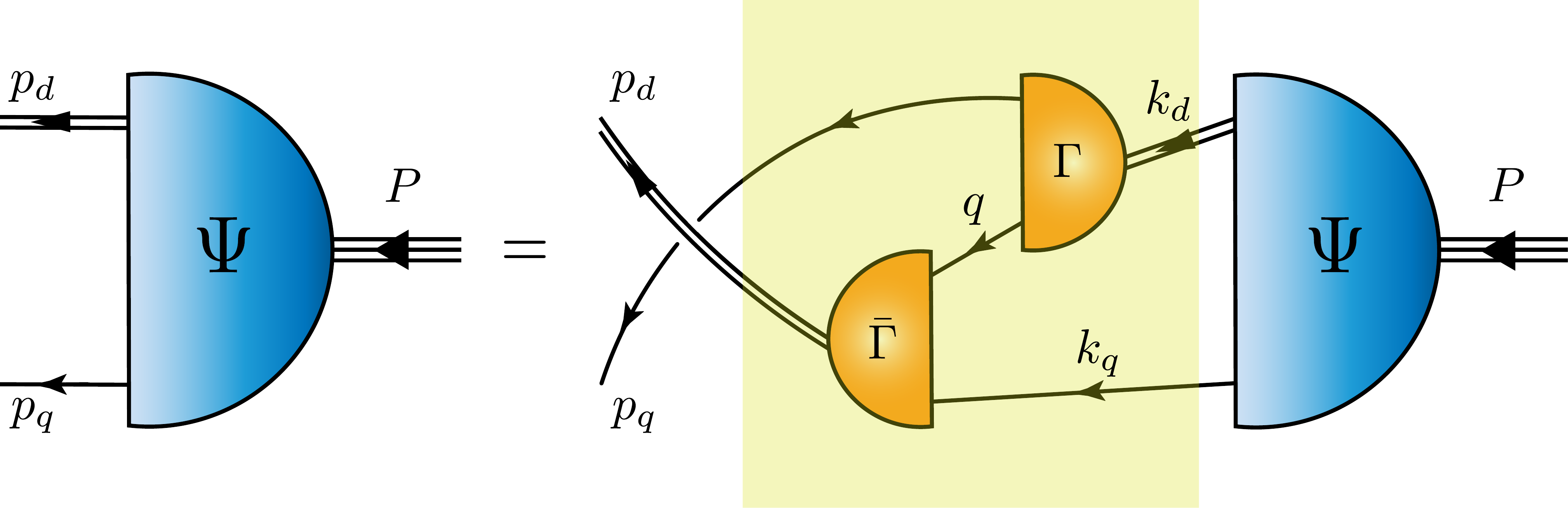}
\end{center}
\vspace*{-0.20cm}
\caption{\label{fig:Faddeev} Poincar\'e-covariant Faddeev equation: a homogeneous linear integral equation for the matrix-valued function $\Psi$, being the Faddeev amplitude for a baryon of total momentum $P = p_q + p_d$, which expresses the relative momentum correlation between the dressed-quarks and -diquarks within the baryon. The (purple) highlighted rectangle demarcates the kernel of the Faddeev equation: {\it single line}, dressed-quark propagator; {\it double line}, diquark propagator; and $\Gamma$, diquark correlation amplitude.
}
\end{figure}

\begin{figure}[!t]
\begin{center}
\includegraphics[width=0.45\textwidth, height=0.25\textheight] {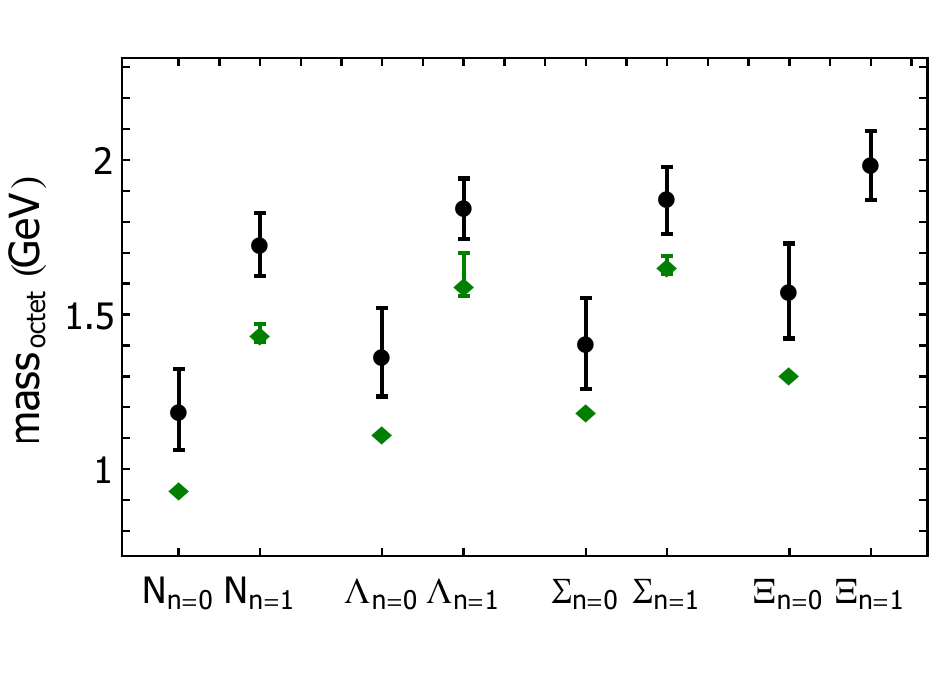}
\hspace*{0.25cm}
\includegraphics[width=0.45\textwidth, height=0.25\textheight] {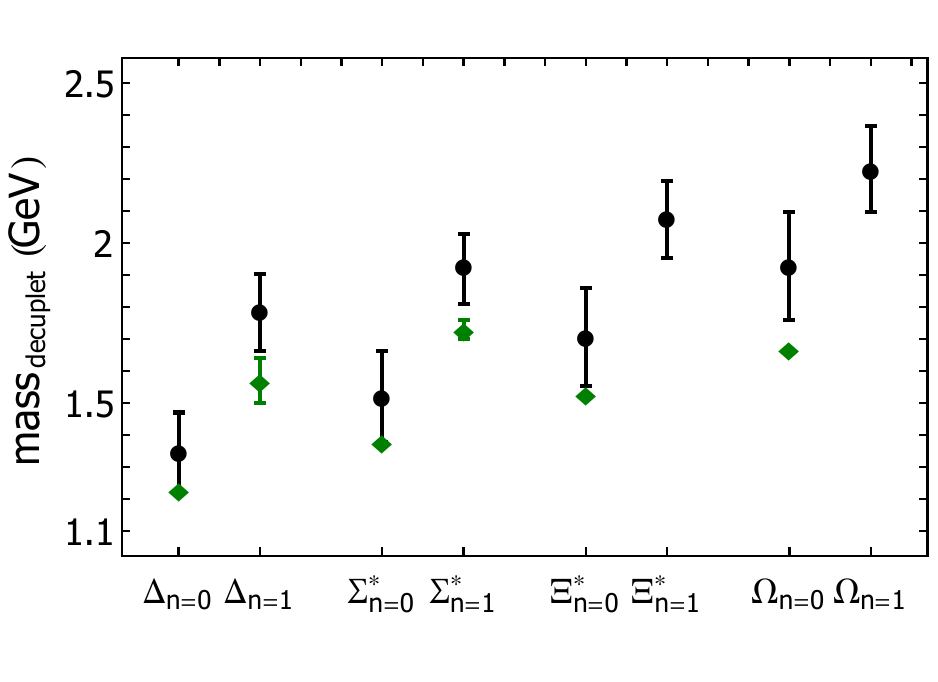}
\end{center}
\vspace*{-0.50cm}
\caption{\label{fig:Spectrum} (Black) solid points are the computed masses for the ground- and first-excited state of the octet (left-panel) and decuplet (right-panel) baryons~\cite{Chen:2019fzn}. The vertical riser indicates the response of our predictions to a coherent $\pm 5$\% change in the mass-scales that appear in the baryon bound-state equation. The horizontal axis lists a particle name with a subscript that indicates whether it is ground-state ($n=0$) or first positive-parity excitation ($n=1$). (Green) diamonds are empirical Breit-Wigner masses take from Ref.~\cite{ParticleDataGroup:2022pth}, the estimated uncertainty in the location of a resonance's Breit-Wigner mass is indicated by an error bar.}
\end{figure}

\begin{figure}[!t]
\begin{center}
\includegraphics[width=0.90\textwidth, height=0.27\textheight] {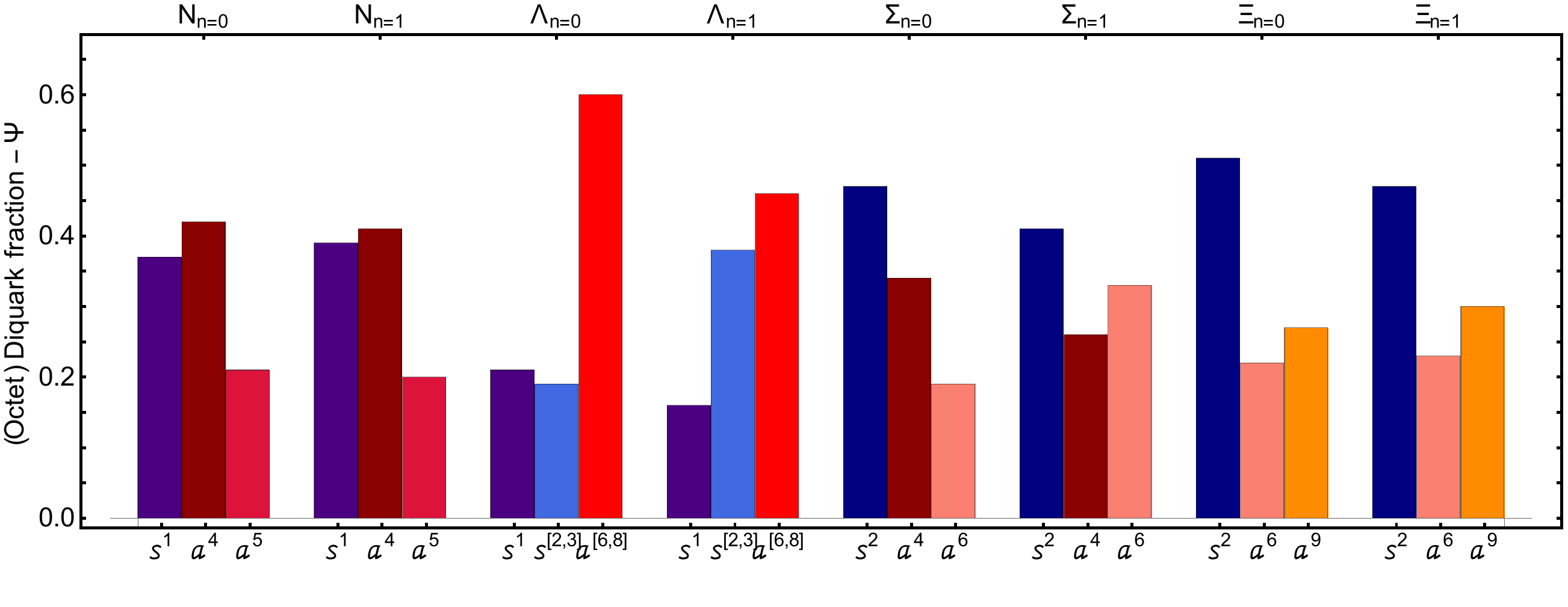}
\includegraphics[width=0.90\textwidth, height=0.27\textheight] {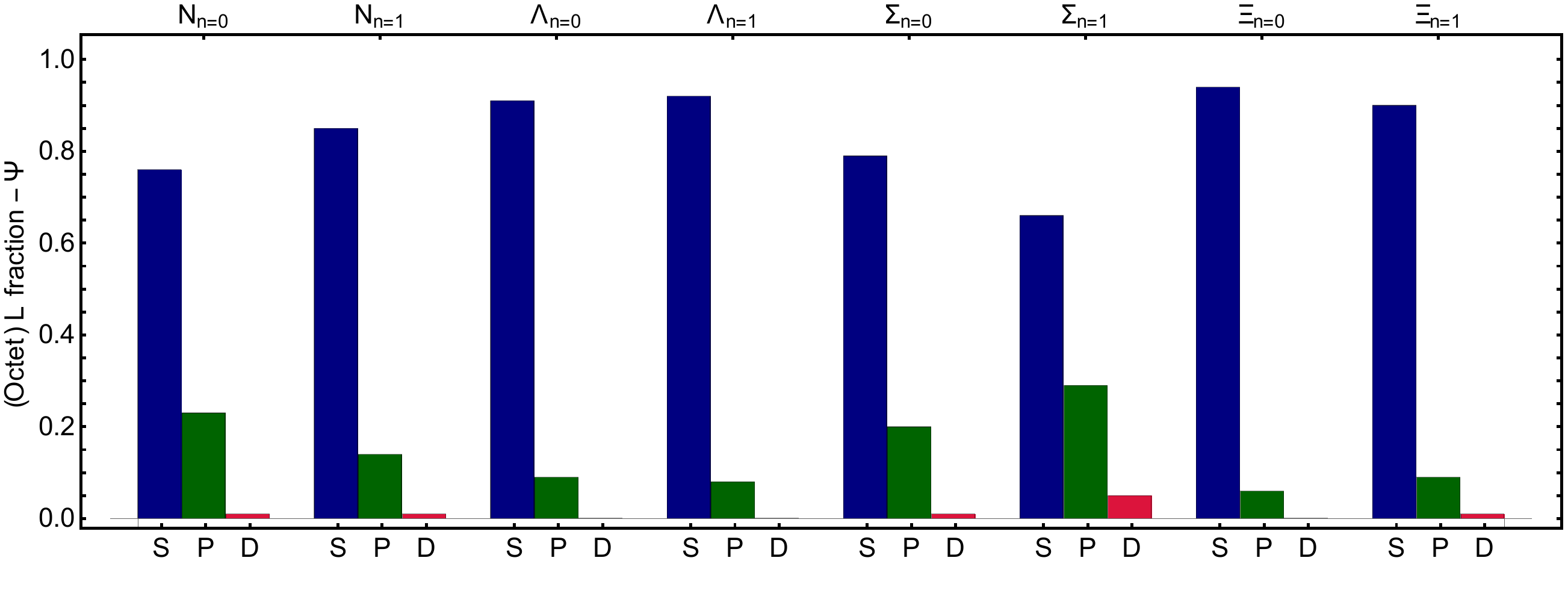}
\includegraphics[width=0.90\textwidth, height=0.27\textheight] 
{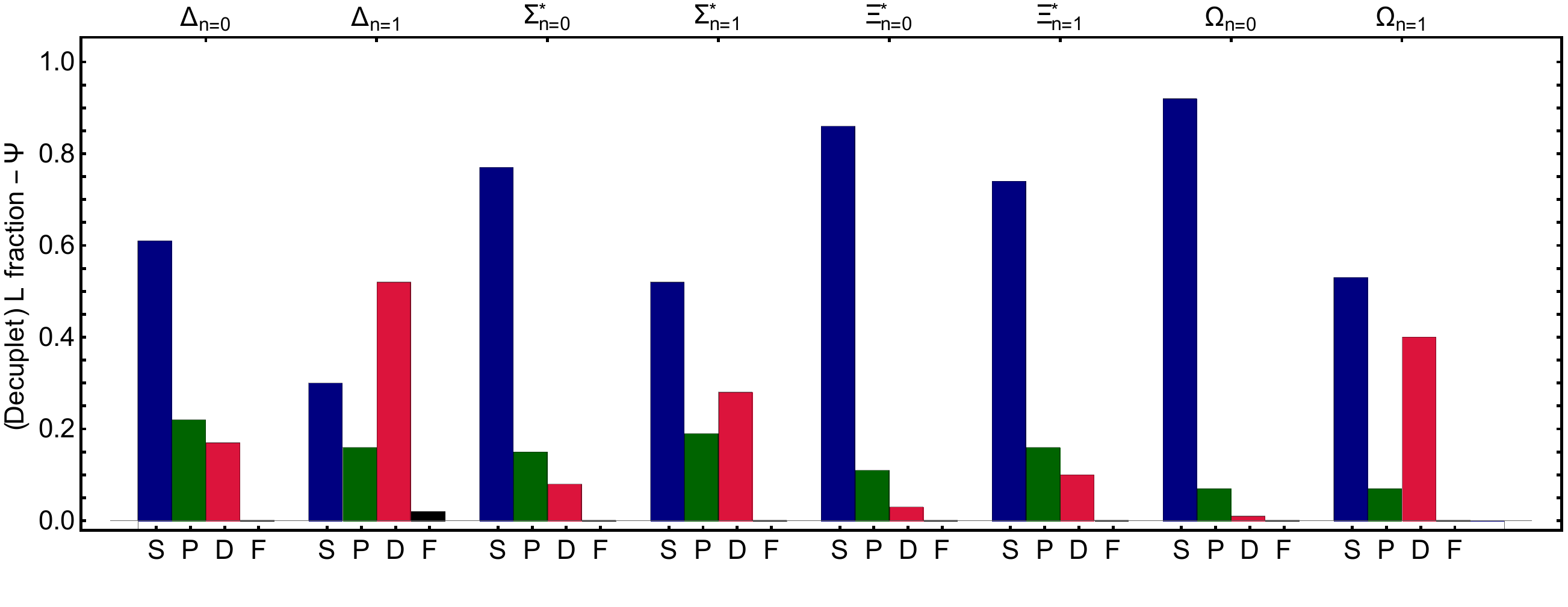}
\end{center}
\vspace*{-0.50cm}
\caption{\label{fig:Structure} Figures adapted from Ref.~\cite{Chen:2019fzn}.
\emph{Top panel:} Relative strengths of various diquark components within the indicated baryon's Faddeev amplitude.
\emph{Middle panel:} Baryon rest-frame quark-diquark orbital angular momentum fractions of ground and first-excited octet baryons.
\emph{Bottom panel:} Baryon rest-frame quark-diquark orbital angular momentum fractions of ground and first-excited octet baryons.
}
\end{figure}

\section{Baryon bound state problem}

The problem of solving the Poincar\'e-covariant Faddeev equation can be transformed into that of solving the linear, homogeneous matrix equation depicted in Fig.~\ref{fig:Faddeev}. This is because two decades of studying baryons as three-quark bound-states~\cite{Oettel:1998bk, Eichmann:2016yit, Lu:2017cln, Yin:2019bxe, Yin:2021uom} have demonstrated the appearance of soft (non-pointlike) fully-interacting diquark correlations within baryons, whose characteristics are greatly influenced by DCSB~\cite{Barabanov:2020jvn}. Note that a baryon described by Fig.~\ref{fig:Faddeev} can be interpreted as a Borromean bound-state where the binding energy is given by two main contributions~\cite{Segovia:2015ufa, Mezrag:2017znp}: One part is expressed in the formation of tight diquark correlations, the second one is generated by the quark exchange depicted in the highlighted rectangle of the Fig.~\ref{fig:Faddeev}. This exchange ensures that no quark holds a special place because each one participates in all diquarks to the fullest extent allowed by the baryon's quantum numbers. The continual rearrangement of the quarks also guarantees that the wave function complies with the fermionic nature of a baryon.

Figure~\ref{fig:Spectrum} shows the computed masses of octet and decuplet baryons and their first positive-parity excitations~\cite{Chen:2019fzn}. It is apparent that the theoretical values are uniformly larger than the corresponding empirical ones. This is because our results should be viewed as those of a given baryon's dressed-quark core, whereas the empirical values include all contributions, including meson-baryon final-state interactions (MB\,FSIs), which typically generate a measurable reduction~\cite{Suzuki:2009nj}. This was explained and illustrated in a study of the nucleon, its parity-partner and their radial excitations~\cite{Chen:2017pse}; and has also been demonstrated using a symmetry-preserving treatment of a vector$\,\times\,$vector contact interaction \cite{Lu:2017cln, Yin:2019bxe, Yin:2021uom}. Identifying the difference between our predictions and experiment as the result of MB\,FSIs, then one finds that such effects are fairly homogeneous across the spectrum. Namely, they act to reduce the mass of ground-state octet and decuplet baryons and their first positive-parity excitations by roughly $0.23(6)\,$GeV.

It is worth noting the emergence of a $\Sigma-\Lambda$ mass-splitting despite the fact that we have assumed isospin symmetry, \emph{i.e.} mass-degenerate $u$- and $d$-quarks, described by the same propagator, so that all diquarks in an isospin multiplet are degenerate. Whilst the $\Lambda^0$ and $\Sigma^0$ baryons are associated with the same combination of valence-quarks, their spin-flavor wave functions are different: the $\Lambda^0_{I=0}$ contains more of the lighter $J=0$ diquark correlations than the $\Sigma^0_{I=1}$. It follows that the $\Lambda^0$ must be lighter than the $\Sigma^0$. Therefore, our result agrees with the prediction of the Gell-Mann–Okubo mass formula~\cite{Gell-Mann:1961omu, Okubo:1961jc, Okubo:1962zzc}.

It is interesting now to dissect the baryon's wave function in various ways and thereby sketch the character of the quark cores that constitute the ground-state octet and decuplet baryons, and their first positive-parity excitations. We begin by exposing their diquark content. Isoscalar-scalar and isovector-pseudovector diquarks are the only ones needed to describe the octet and decuplet baryons. Since the $I=3/2$ baryons has only isovector-pseudovector diquark content, top panel of Fig.~\ref{fig:Structure} shows the relative size of each diquark contribution to the wave function of octet baryons. As one can see, both scalar and axial-vector diquarks are important in all cases and thus the suppression of one of them makes the calculation unrealistic. Moreover, the nucleon and its first positive-parity excitation possess very similar diquark content.

Middle and bottom panels of Fig.~\ref{fig:Structure} expose the rest-frame orbital angular momentum content of octet and decuplet baryons, respectively. Plainly, every one of the systems considered is primarily $S$-wave in nature, since they are not generated by the Faddeev equation unless $S$-wave components are contained in the wave function. Besides, $P$-wave components play a measurable role in octet ground-states and their first positive-parity excitations; they are attractive in ground-states and repulsive in the excitations. Regarding decuplet systems, $P$-wave components generate a little repulsion, some notable attraction is provided by $D$-waves, and $F$-waves have no measurable impact.

It is important to highlight that a possible way to evade the effects of MB\,FSIs as well as to assess the impact of various diquark and rest-frame orbital angular momentum components in the baryon's wave function is studying electromagnetic elastic and transition form factors of nucleon and $\Delta$ resonances~\cite{Segovia:2013uga, Segovia:2013rca, Segovia:2014aza, Segovia:2015hra, Segovia:2016zyc, Lu:2019bjs, Raya:2021pyr}. A few of which will be sketched below.


\section{The $\mathbf{\gamma^{\ast}p\to N(940),\,N(1440)}$ Transitions}

We are going to review herein the calculation of the nucleon's elastic form factors and the so-called equivalent Dirac and Pauli form factors of the $\gamma^{\ast}N(940)\to N(1440)$ reaction. This section is mostly based on the work presented in Refs.~\cite{Chen:2018nsg, Cui:2020rmu}.

The computation of the desired elastic and transition form factors is a straightforward numerical exercise once the Faddeev amplitudes for the participating states are in hand and the electromagnetic current is specified.  When the initial and final states are $I=1/2$, $J=1/2^+$ baryons, the current is completely determined by two form factors, \emph{viz}.
\begin{equation}
\bar u_{f}(P_f)\big[ \gamma_\mu^T F_{1}^{fi}(Q^2)+\frac{1}{m_{{fi}}} \sigma_{\mu\nu} Q_\nu F_{2}^{fi}(Q^2)\big] u_{i}(P_i)\,,
\label{NRcurrents}
\end{equation}
where: $u_{i}$, $\bar u_{f}$ are, respectively, Dirac spinors describing the incoming/outgoing baryons, with four-momenta $P_{i,f}$ and masses $m_{i,f}$ so that $P_{i,f}^2=-m_{i,f}^2$; $Q=P_f-P_i$; $m_{{fi}} = (m_f+m_{i})$; and $\gamma^T \cdot Q= 0$.

The vertex sufficient to express the interaction of a photon with a baryon generated by the Faddeev equation in Fig.~\ref{fig:Faddeev} is described elsewhere~\cite{Segovia:2014aza}. It is a sum of six terms, depicted in the Appendix~C of Ref.~\cite{Segovia:2014aza}, with the photon probing separately the quarks and diquarks in various ways, so that diverse features of quark dressing and the quark-quark correlations all play a role in determining the form factors.  


\begin{figure}[!t]
\begin{center}
\includegraphics[width=0.45\textwidth, height=0.25\textheight]{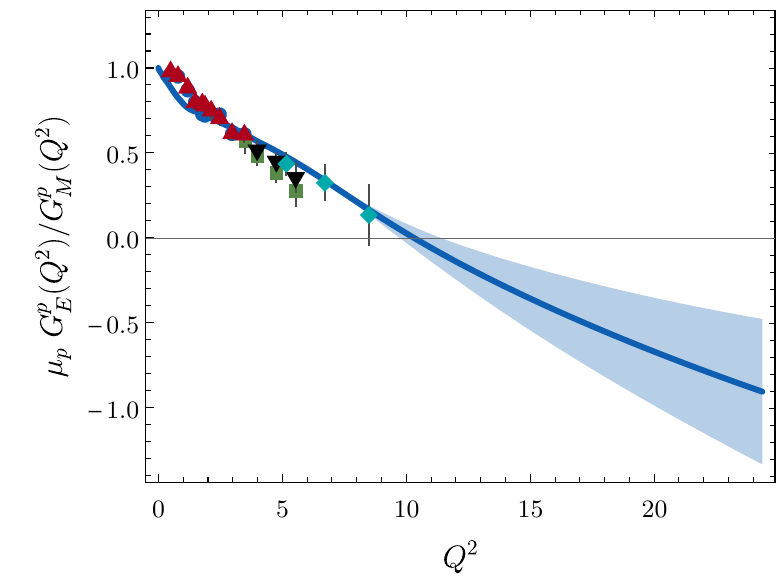}
\hspace*{0.25cm}
\includegraphics[width=0.45\textwidth, height=0.25\textheight]{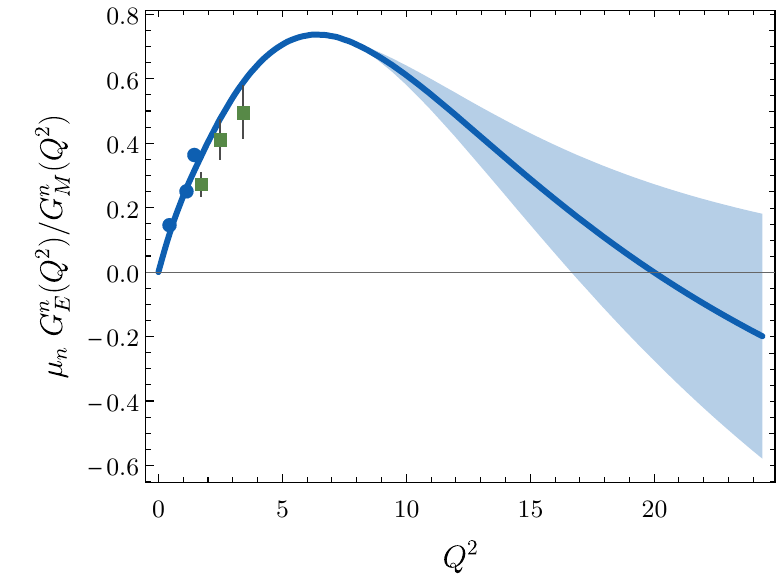}
\end{center}
\vspace*{-0.50cm}
\caption{\label{fig:NucNuc} Ratios of Sachs form factors, $\mu_N G_E^N(x)/G_M^N(x)$.
\emph{Left panel:}  Proton case compared with data (red up-triangles~\cite{JeffersonLabHallA:1999epl}; green squares~\cite{JeffersonLabHallA:2001qqe}; blue circles~\cite{Punjabi:2005wq}; black down-triangles~\cite{Puckett:2010ac}; and cyan diamonds~\cite{Puckett:2011xg}). \emph{Right panel:} Neutron case compared with data (blue circles~\cite{E93-038:2003ixb} and green squares~\cite{Riordan:2010id}).}
\end{figure}

Data on $R_{EM}^p(Q^2)=\mu_p G_E^p(Q^2)/G_M^p(Q^2)$, obtained using polarisation transfer reactions at JLab, show a trend toward zero with increasing momentum-transfer-squared~\cite{JeffersonLabHallA:1999epl, JeffersonLabHallA:2001qqe, Punjabi:2005wq, Puckett:2010ac, Puckett:2011xg}. We depict this ratio and its analogue for the neutron in Fig.~\ref{fig:NucNuc}. As shown in Ref.\,\cite[Fig.\,4]{Segovia:2015ufa}, these form factors are very sensitive to those scalar-diquark components of the nucleon's rest-frame Faddeev wave function that carry nonzero quark+diquark orbital angular momentum.  Consequently, the appearance and location of a zero in $R_{EM}^N(Q^2)$ measures the strength of both quark-quark and angular momentum correlations within the nucleon.  Both are expressions of EHM.

\begin{figure}[!t]
\hspace*{-0.40cm}
\includegraphics[clip, width=0.45\textwidth, height=0.25\textheight]{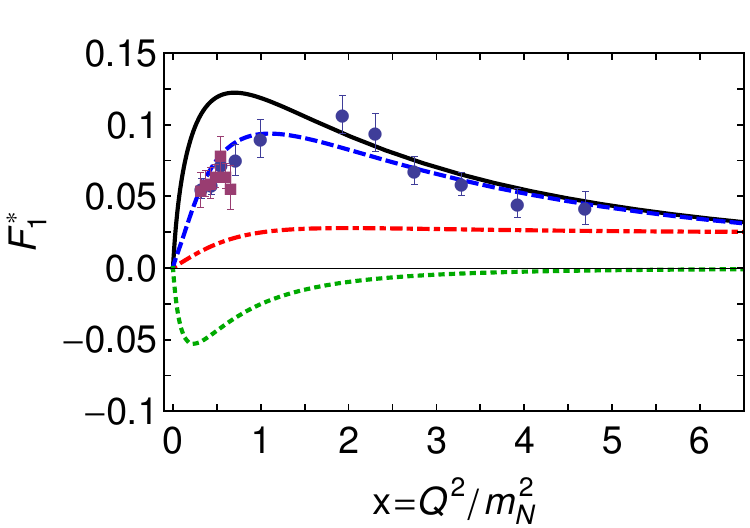}
\hspace*{0.25cm}
\includegraphics[clip, width=0.45\textwidth, height=0.25\textheight]
{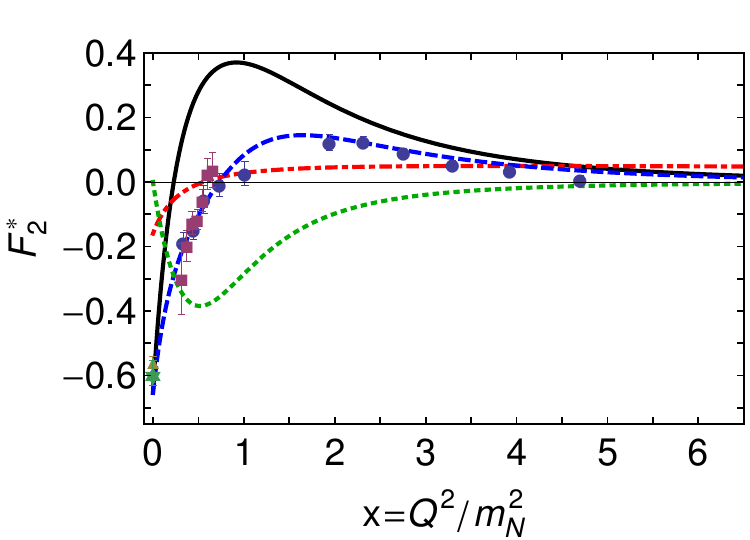}
\put(-60,82){\small QCD-kindred}
\put(-60,70){\small \textcolor{red}{CI-model}}
\put(-60,58){\small \textcolor{blue}{Fit}}
\put(-60,46){\small \textcolor{dgreen}{MB-FSIs}}  \\
\includegraphics[clip, width=0.45\textwidth, height=0.25\textheight]{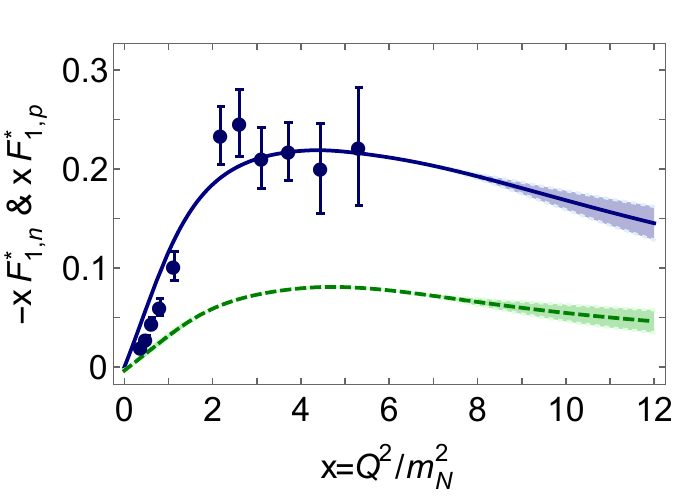}
\hspace*{0.25cm}
\includegraphics[clip, width=0.45\textwidth, height=0.25\textheight]{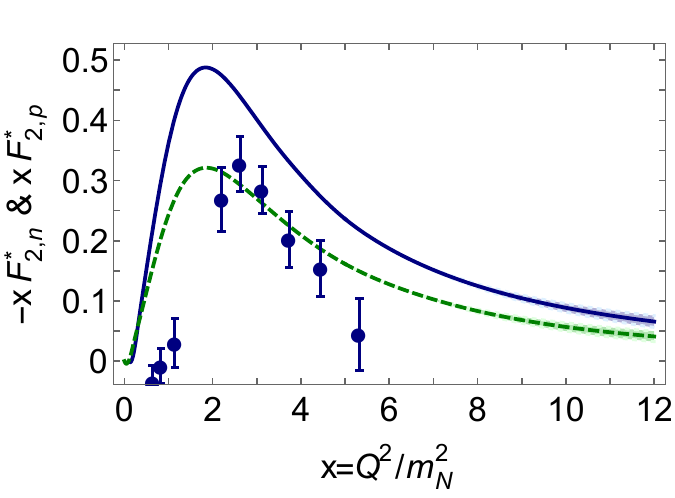}
\caption{\label{fig:NucRop} \emph{Top-left:} Dirac transition form factor, $F_{1}^{\ast}(x)$, $x=Q^2/m_N^2$. Solid (black) curve, QCD-kindred prediction; dot-dashed (red) curve, contact-interaction result; dotted (green) curve, inferred meson-cloud contribution; and dashed (blue) curve, anticipated complete result. 
\emph{Top-right:} Pauli transition form factor, $F_{2}^{\ast}(x)$, with same legend as top-left panel. Data in both panels: circles (blue)~\cite{CLAS:2009ces}; triangle (gold)~\cite{CLAS:2009tyz}; squares (purple)~\cite{CLAS:2012wxw}; and star (green)~\cite{ParticleDataGroup:2022pth}.
\emph{Bottom-left:} Computed $x$-weighted Dirac transition form factor for the reactions $\gamma^\ast\,p\to R^+$ (solid blue curves) and $\gamma^\ast\,n\to R^0$ (dashed green curves). In all cases, the results on $x\in [6,12]$ are projections, obtained via extrapolation of analytic approximations to our results on $x\in [0,6]$ (see Ref.~\cite{Chen:2018nsg} for details). The width of the band associated with a given curve indicates our confidence in the extrapolated value. Data in both panels are for the charged channel transitions, $F_{1,p}^\ast$ and $F_{2,p}^\ast$: circles (blue)~\cite{CLAS:2009ces}. No data currently exist for the neutral channel but they are expected.
\emph{Bottom-right:} Computed $x$-weighted Pauli transition form factor, with same legend as bottom-left panel.
}
\vspace*{-0.40cm}
\end{figure}

The equivalent Dirac and Pauli form factors of the $\gamma^{\ast}p\to R^+$ transition are displayed in the top row of Fig.~\ref{fig:NucRop}. The results obtained using QCD-derived propagators and vertices agree with the data on $x=Q^2/m_N^2\gtrsim 2$. The contact-interaction result~\cite{Wilson:2011aa} simply disagrees both quantitatively and qualitatively with the data. Therefore, experiment is evidently a sensitive tool with which to chart the nature of the quark-quark interaction and hence discriminate between competing theoretical hypotheses. The disagreement between the QCD-kindred result and data on $x\lesssim 2$ is due to meson-cloud contributions that are expected to be important on this domain. An inferred form of that contribution is provided by the dotted (green) curves in top-left and -right panels of Fig.~\ref{fig:NucRop}. They are small already at $x=2$ and vanish rapidly thereafter so that the quark-core prediction remain as the explanation of the data. It is worth to emphasize that the zero crossing in $F_{2}^{\ast}$ is always present but its precise location depends on the meson-cloud estimation.

Finally, since it is anticipated that CLAS~12 detector will deliver data on the Roper-resonance electro-production form factors out to $Q^2 \sim 12 m_N^2$, we depict in the bottom row of Fig.~\ref{fig:NucRop} the $x$-weighted Dirac and Pauli transition form factors for the reactions $\gamma^\ast p \to R^{+}$, $\gamma^\ast n\to R^{0}$ on the domain $0<x<12$. As one can see, there is no indication of the scaling behaviour expected of the transition form factors: $F^\ast_{1} \sim 1/x^2$, $F^\ast_2 \sim 1/x^3$. Since each dressed-quark in the baryons must roughly share the impulse momentum, $Q$, we expect that such behavior will only become evident on $x\gtrsim 20$.

\section{Summary}

We have highlighted the importance of the emergent hadron mass mechanism and its role in generating the visible mass of the universe. The review includes an analysis of gluon and quark propagators as well as the effective color strength between quarks, antiquarks and gluons. Therewith, we have described baryon bound states within the named quark-diquark picture of the Poincar\'e-covariant Faddeev equation, where diquark correlations are produced due to the peculiarities of the strong force. In identifying their observable consequences, we have computed the lowest-lying octet and decuplet baryons, offering a comprehensive understanding of their internal structure trough their diquark content and rest-frame orbital angular momentum components; finally, the electromagnetic form factors of the nucleon and its first radial excitation have been shown as examples to avoid MB\,FSIs soft contributions and illuminate the so-called dressed-quark core of nucleon resonances.


\acknowledgments
Work partially financed by Ministerio Espa\~nol de Ciencia e Innovaci\'on under grant No. PID2022-140440NB-C22; Junta de Andaluc\'\i a under contract Nos. PAIDI FQM-370 and PCI+D+i under the title: ``Tecnolog\'\i as avanzadas para la exploraci\'on del universo y sus componentes" (Code AST22-0001).


\bibliographystyle{JHEP}
\bibliography{QNP2024_JS_CSM-Diquarks}

\end{document}